\documentclass[sigconf,screen]{acmart}

\usepackage{balance}

\setcopyright{none}
\copyrightyear{}
\acmYear{}
\acmDOI{}
\acmISBN{}
\acmConference[CHI'24 Workshop on Future of Cognitive Personal Informatics]{CHI'24 Workshop on Future of Cognitive Personal Informatics}{May 11--16, 2024}{Honolulu, HI, USA}
\acmBooktitle{CHI'24 Workshop on Future of Cognitive Personal Informatics}

\begin{document}

\title{Modeling User Preferences via Brain-Computer Interfacing}

\author{Luis A. Leiva}
\orcid{0000-0002-5011-1847}
\affiliation{
  \institution{University of Luxembourg}
  \country{Luxembourg}
}

\author{V. Javier Traver}
\orcid{0000-0002-1596-8466}
\affiliation{
  \institution{INIT, Universitat Jaume I}
  \country{Spain}
}

\author{Aleksandra Kawala-Sterniuk}
\orcid{0000-0001-7826-1292}
\affiliation{
  \institution{Opole University of Technology}
  \country{Poland}
}

\author{Tuukka Ruotsalo}
\orcid{0000-0002-2203-4928}
\affiliation{
  \institution{University of Copenhagen}
  \country{Denmark}
}
\affiliation{
  \institution{LUT University}
  \country{Finland}
}

\renewcommand{\shortauthors}{L. A. Leiva, V. J. Traver, A. Kawala-Sterniuk, T. Ruotsalo}

\begin{abstract}
Present Brain-Computer Interfacing (BCI) technology allows inference and detection of cognitive and affective states, but fairly little has been done to study scenarios in which such information can facilitate new applications that rely on modeling human cognition. One state that can be quantified from various physiological signals is attention. Estimates of human attention can be used to reveal preferences and novel dimensions of user experience. Previous approaches have tackled these incredibly challenging tasks using a variety of behavioral signals, from dwell-time to clickthrough data, and computational models of visual correspondence to these behavioral signals. However, behavioral signals are only rough estimations of the real underlying attention and affective preferences of the users. Indeed, users may attend to some content simply because it is salient, but not because it is really interesting, or simply because it is outrageous. With this paper, we put forward a research agenda and example work using BCI to infer users’ preferences, their attentional correlates towards visual content, and their associations with affective experience. Subsequently, we link these to relevant applications, such as information retrieval, personalized steering of generative models, and crowdsourcing population estimates of affective experiences.
\end{abstract}



\maketitle

\section{Introduction}

Estimating which parts of some digital content are likely to draw the users’ interest, and whether the parts are experienced with high or low intensity, positively, or negatively by the user. However, classification of cognitive and affective states provide only simple estimates of how neural responses map to discrete states of human cognition. Conversely, recent research has recognized that estimating cognitive and affective states, such as attention, valence, and arousal can be effectively used for many downstream tasks even when the accuracy of the upstream classification model is modest~\cite{Davis_2022_CVPR,9353984,de2020brain,brainsource:2020}. These approaches build upon developing models of personal cognitive profiles or crowd models that can be effective even when single-trial classification cannot reach robust performance. We demonstrate the use of such models in various downstream applications, including information retrieval, affective similarity estimation, steering generative models, and crowdsourced approaches.

Further benefits of attention and affective estimation have been demonstrated in crowd settings, in which reactions of many individuals can be combined for more consistent and reliable estimates of what in content drags users' attention and how they experience that content. 
Here, we put forward a research agenda to study inference of human attention and preferences from BCI as captured in response to naturalistic perception of digital information. Furthermore, we present how sigals from crowds of users can be used via \emph{brainsourcing}, i.e. crowdsourced BCI signal acquisition~\cite{brainsource:2020}. This can allow an accurate estimation of user preferences, attention allocation, and---critically---the affective component of attention, directly measured from the natural and implicit brain potentials evoked in users' responses. We demonstrate how to gather and utilize the resulting data in single-user and crowdsourcing settings to reveal how users react to different stimuli and how their attention and affective responses can be predicted. These responses produce consistent measures of user experiences that go beyond of what is possible with behavioral data.


\section{Research Challenges}

In the following we summarize current research challenges and example works in BCI instrumentation, preference estimation, predicting affective states, crowdsourcing affective annotation, and the importance of open data.

\subsection{Instrumentation} 


Both electroencephalography (EEG) or functional Near-Infrared Spectroscopy (fNIRS)
stand out as the predominant non-invasive methods for acquiring brain data~\cite{kawala2021summary}. 
EEG directly captures the brain's bio-electrical activity by recording electrical fluctuations via electrodes placed on the scalp~\cite{kawala2021summary}. In contrast, fNIRS relies on optical techniques to detect changes in hemodynamics~\cite{wieczorek2023custom}, typically induced by cortical responses during motor, cognitive, and perceptual functions of the brain. Each method has its own set of advantages and disadvantages, often leading to their integration for a more comprehensive examination~\cite{pelc2023machine}.


EEG signals, for example, have usually a low signal-to-noise ratio and low spatial resolution~\cite{beres2017}. 
This makes it difficult to identify what brain areas are activated by a particular response~\cite{Srinivasan99}.
In fNIRS, \emph{scattering}, occurring about {$100$} times more frequently than \emph{absorption}, leads to light attenuation. 
The longer the path of the photon due to scattering, the greater the likelihood of absorption. 
However, the light emitted by a source can be captured by multiple detectors, 
eliminating thus the necessity for additional components~\cite{wieczorek2023custom}. 


\subsection{Estimating Preferences} 

One of the lasting challenges for user modelling has been to recognize and predict individual preferences. 
Recently, neurophysiological data has been employed for learning user prefences. 
The advantage of such data is that it can be obtained implicitly without requiring any intentional interaction. 
As such, preferences can be inferred from brain data captured in response to natural stimuli users are experiencing. 
However, the reliability of such information and its utility in downstream applications has not been studied. 
Our previous work has proved useful for quantifying the reliability of preferences inferred from brain responses 
and their utility as feedback to generative models to produce new, unseen information that matches the inferred preference information~\cite{9987515, 9353984}.


Recently, it has become possible to predict preference information from brain signals measured via EEG and fNIRS. 
We studied whether individuals' preferences contradict with group preferences~\cite{9987515}. 
That is, whether we can detect preferences of individuals toward group preferences 
even when the individual does not explicitly disclose such preference. 
Experimental evidence shows that brain activity collected from participants produced in response to viewing images 
is associated with their self-reported preferences. 
Our results show that brain responses present a graded response to preferences, 
and that brain responses alone can be used to train classifiers that reliably estimate preferences. 
Furthermore, we show that brain responses reveal additional preference information that correlates with group preference, 
even when participants self-reported having no such preference. 
Our analysis of brain responses carries significant implications for using brain responses for preference inference, 
as it suggests an individual's explicitly reported preferences are not always aligned with the preferences inferred from their brain responses. 
These findings call into question the reliability of explicit and behavioral signals. 
They also imply that additional, multimodal sources of information may be necessary to infer reliable preference information.


Preference information can also be used in downstream applications to adapt content and user interfaces according to the estimated preferences. 
We have studied such cognitive integration with generative image models to model personal attraction~\cite{9353984}. 
We demonstrate models that use generative adversarial neural networks (GANs) 
to model subjective preferences unconstrained by predefined model parameterization. 
GANs are coupled with brain-computer interfaces to capture personalized attractiveness reactions toward images generated from a model. 
These reactions are then used to control a GAN model, 
finding a representation that matches the features constituting an attractive image for an individual. 
We show that our approach yielded highly accurate generative outputs and replicate findings from social neuroscience, 
suggesting that the individually responsive, generative nature of GANs and BCI provide a powerful, new tool 
in mapping individual differences and visualizing cognitive-affective processing.

\subsection{Relevance and Affective States} 


Information retrieval (IR) relies on a general notion of relevance, 
which is used as the principal foundation for ranking and evaluating methods. 
However, IR does not account for more a nuanced affective experiences of users.
We consider the emotional response decoded directly from the human brain as an alternative dimension of relevance~\cite{Ruotsalo23a_affective}.
We report an experiment covering seven different scenarios 
in which we measure and predict how users emotionally respond to visual image contents 
by using fNIRS neuroimaging on two commonly used affective dimensions: 
\emph{valence} (negativity and positivity) and \emph{arousal} (boredness and excitedness). 
Our results show that affective states can be successfully decoded using fNIRS,
and utilized to complement the present notion of relevance in IR studies. 
Our work opens new avenues for incorporating emotional states in IR evaluation, affective feedback, and information filtering.


Affective information extends the notion of relevance toward understanding experiences of visual similarity. 
The present notion of visual similarity is based on features derived from image contents. 
This ignores the users' emotional or affective experiences toward the content, and how users feel when they search for images. 
We consider valence, a positive or negative quantification of affective appraisal, as a novel dimension of image similarity. 
We report the largest neuroimaging experiment that quantifies and predicts the valence of visual content by using functional near-infrared spectroscopy from brain-computer interfacing~\cite{Ruotsalo23b_affective}. 
We show that affective similarity can be (1)~decoded directly from brain signals in response to visual stimuli, 
(2)~utilized for predicting affective image similarity with an average accuracy of 0.58 and an accuracy of 0.65 for high-arousal stimuli, 
and (3)~effectively used to complement affective similarity estimates of content-based models; 
for example when fused fNIRS and image rankings the retrieval F-measure@20 is 0.70. 
This work encourages new research on affective multimedia analysis, retrieval, and user modeling.

\subsection{Crowdsourcing Affective Annotations} 


Automatic annotation of multimedia contents in terms of their affective contents can be useful for a range of interactive tasks. 
A convenient alternative to content-based analysis is to rely on brain signals from several participants who are exposed to those contents. 
In this line of research, a brainsourcing experiment was conducted~\cite{Ruotsalo23tac} 
relying on the fNIRS signals from 31 participants who just passively watched a set of images. 
It was shown that the prediction of the valence and arousal of the images improves with the crowdsize. 
For example, the mean accuracy of individual predictions for valence is lower than 60\%, and increases up to 65\% with a crowdsize of just 5 participants. 


More recently, the concept has been explored also in the context of EEG signals and video stimuli, with similarly positive results~\cite{Moreno24paa}. 
In this case, however, it was observed that the crowdsourcing was not effective for all videos. 
Therefore, an open issue is to either improve the prediction to be more generally effective, 
or to identify under which conditions the brainsourcing-based prediction is not reliable. 
The source of the difficulty may be in either the source stimuli (whose corresponding emotions might be more ambiguous), 
or in the quality of the brain signals, or in the limitations of the predictive machine learning model or the robustness of the agreement procedure.

\subsection{Importance of Open Data} 


A critical aspect to boost the research in the scope of affective BCI is high-quality public datasets, 
well-documented and with accompanying code, processed data, and optionally pre-trained models, 
following the FAIR principles\footnote{\url{https://www.go-fair.org/fair-principles}}. 
In this regard, the NEMO dataset~\cite{Spape23tac} has been recently released, 
featuring fNIRS recordings from thirty-one participants who engaged in an emotional perception task and in affective imagery task.


An equally important ingredient is sticking to good practices in terms of data usage and reporting. 
An obvious but important example is avoiding data leakage. 
For example, in the context of brain signals it is relatively common to split the EEG signal into temporal segments 
which are treated as independent instances. 
Although this increases the number of training/testing instances, 
segments temporally close in the signal 
or coming from the same participant and/or the same eliciting stimulus can be highly correlated~\cite{9264220,ahmed2021confounds}. 
Furthermore, in supervised learning, these segments usually inherit the sequence-level label, 
which might not always be a good choice since different temporal parts of the brain signal may carry different information,
mostly for long stimulus such as video snippets. 
Thus, generally speaking, depending on how data is partitioned into training and test sets, the performance can vary notably~\cite{Yoelvis24bmel}, 
which may hinder the advances in the field in various ways. 
In the deep learning area, with complex models which can easily overfit the training set, 
it can be hard to find out whether the reported high performances are actually attributable 
to the merits of the proposed algorithms and models or they implicitly hide data-related artifacts.  

Two main challenges remain for BCI-related datasets. 
On the one hand, large-scale datasets are required to properly train the data-hungry deep learning models. 
This is currently difficult in the context of brain signals, due to the cost of the equipment 
and the huge effort involved in long recording sessions with multiple participants. 
In the future, one may envision foundation models in the context of BCI, with BENDR being one example~\cite{Kostas21fhn}. 
On the other hand, multimodality (e.g. including several types of brain signals, eye tracking data, 
diverse multimedia stimuli beyond images/sound/video) would facilitate research on a wider and richer set of topics 
and more powerful models, which can more heavily rely on self-supervised learning, and explore joint and multi-task learning.

\subsection{Ethical Implications}

Portable sensor technology has already made possible real-time monitoring of physiological signals. 
This opens new avenues for individuals and data-based service providers. 
However, it also introduces risks for privacy and user autonomy~\cite{mizgajski2019affective}.  
As user monitoring technology evolves to identify emotional characteristics, 
reaching a level of prevalence similar to behavioral tracking on personal computers and smartphones, 
service providers may gain access to affective information. 
Consequently, the potential for unethical applications of physiological monitoring 
and other wearable hardware to expose cognitive and emotional user attributes may arise. 
Recent studies suggest that growing awareness of data usage is causing users to exhibit increased caution 
when engaging with technology capable of divulging detailed information about their affective and cognitive states~\cite{prange2021investigating}. 

The capacity to model human attention, affect, preferences and behaviors 
with unprecedented details has already transformed the Internet and service economy, 
but physiological data can provide an additional layer of rich, personal, 
and sensitive data that can enhance user experiences and provide improved benefits to users. 
On the other hand, neurophysiological data are multifaceted and can be utilized for unintended use. 
This makes it comparable to the privacy risks of those data presently used to model users in their digital environments. 
However, neurophysiological data, while noisy, has much higher fidelity in terms of which dimensions of users' experiences it allows to be inferred. 
As we start to live with sensors capable of measuring neurophysiological signals extensively 
and provide it as input for various applications, as also discussed in this paper, 
it can lead to systems that are much more accurate than what the sensor accuracies reported today would suggest. 

Research and regulatory measures must guarantee that data is utilized solely for purposes explicitly agreed upon by users. 
It is imperative to establish privacy methods and practices empowering users to maintain control over the utilization 
and sharing of the data gathered from them. 
Although it might seem straightforward at first, neurophysiological data enabling affective inference is considerably more sensitive 
than the current behavioral data collected by computing services. 
Therefore, the data must be used ethically and regulated with adequate policies 
that can prevent lasting threats to the public. To this end, there are already regulative actions circumventing unethical use. For instance, the EU AI act\footnote{\url{https://digital-strategy.ec.europa.eu/en/policies/regulatory-framework-ai}} prevents AI systems for the purpose of identifying or inferring emotions or intentions of natural persons on the basis of their biometric data in the workplace and educational institutions.

\section*{Acknowledgments}
\sloppy
Research supported by the Horizon 2020 FET program of the European Union (grant CHIST-ERA-20-BCI-001),
the European Innovation Council Pathfinder program (SYMBIOTIK project, grant 101071147),
and the National Science Centre, Poland (grant 2021/03/Y/ST7/00008).
This publication is part of the project PCI2021-122036-2A, funded by MCIN/AEI/10.13039/5011\linebreak[0]{}00011033 and European Union NextGenerationEU/\linebreak[0]{}PRTR. 
This work also received support from the Academy of Finland (grants 322653, 328875, 336085, 350323, and 352915).
\balance
\bibliographystyle{ACM-Reference-Format}
\bibliography{references}


\begin{thebibliography}{21}


\ifx \showCODEN    \undefined \def \showCODEN     #1{\unskip}     \fi
\ifx \showDOI      \undefined \def \showDOI       #1{#1}\fi
\ifx \showISBNx    \undefined \def \showISBNx     #1{\unskip}     \fi
\ifx \showISBNxiii \undefined \def \showISBNxiii  #1{\unskip}     \fi
\ifx \showISSN     \undefined \def \showISSN      #1{\unskip}     \fi
\ifx \showLCCN     \undefined \def \showLCCN      #1{\unskip}     \fi
\ifx \shownote     \undefined \def \shownote      #1{#1}          \fi
\ifx \showarticletitle \undefined \def \showarticletitle #1{#1}   \fi
\ifx \showURL      \undefined \def \showURL       {\relax}        \fi
\providecommand\bibfield[2]{#2}
\providecommand\bibinfo[2]{#2}
\providecommand\natexlab[1]{#1}
\providecommand\showeprint[2][]{arXiv:#2}

\bibitem[Ahmed et~al\mbox{.}(2021)]%
        {ahmed2021confounds}
\bibfield{author}{\bibinfo{person}{Hamad Ahmed}, \bibinfo{person}{Ronnie~B. Wilbur}, \bibinfo{person}{Hari~M. Bharadwaj}, {and} \bibinfo{person}{Jeffrey~Mark Siskind}.} \bibinfo{year}{2021}\natexlab{}.
\newblock \showarticletitle{Confounds in the data—Comments on “Decoding brain representations by multimodal learning of neural activity and visual features”}.
\newblock \bibinfo{journal}{\emph{IEEE Trans. Pattern Anal. Mach. Intell.}} \bibinfo{volume}{44}, \bibinfo{number}{12} (\bibinfo{year}{2021}), \bibinfo{pages}{9217--9220}.
\newblock
\urldef\tempurl%
\url{https://doi.org/10.1109/TPAMI.2021.3121268}
\showDOI{\tempurl}


\bibitem[Beres(2017)]%
        {beres2017}
\bibfield{author}{\bibinfo{person}{Anna~M. Beres}.} \bibinfo{year}{2017}\natexlab{}.
\newblock \showarticletitle{Time is of the Essence: A Review of Electroencephalography ({EEG}) and Event-Related Brain Potentials ({ERPs}) in Language Research}.
\newblock \bibinfo{journal}{\emph{Applied Psychophysiology and Biofeedback}} \bibinfo{volume}{42}, \bibinfo{number}{4} (\bibinfo{year}{2017}), \bibinfo{pages}{247--255}.
\newblock
\urldef\tempurl%
\url{https://doi.org/10.1007/s10484-017-9371-3}
\showDOI{\tempurl}


\bibitem[Davis et~al\mbox{.}(2022)]%
        {Davis_2022_CVPR}
\bibfield{author}{\bibinfo{person}{Keith~M. Davis}, \bibinfo{person}{Carlos de~la Torre-Ortiz}, {and} \bibinfo{person}{Tuukka Ruotsalo}.} \bibinfo{year}{2022}\natexlab{}.
\newblock \showarticletitle{Brain-Supervised Image Editing}. In \bibinfo{booktitle}{\emph{Proc. IEEE/CVF Conf. on Computer Vision and Pattern Recognition}} \emph{(\bibinfo{series}{CVPR '22})}. \bibinfo{pages}{18480--18489}.
\newblock
\urldef\tempurl%
\url{https://doi.org/10.1109/CVPR52688.2022.01793}
\showDOI{\tempurl}


\bibitem[Davis et~al\mbox{.}(2020)]%
        {brainsource:2020}
\bibfield{author}{\bibinfo{person}{Keith~M. Davis}, \bibinfo{person}{Lauri Kangassalo}, \bibinfo{person}{Michiel Spap\'{e}}, {and} \bibinfo{person}{Tuukka Ruotsalo}.} \bibinfo{year}{2020}\natexlab{}.
\newblock \showarticletitle{Brainsourcing: Crowdsourcing Recognition Tasks via Collaborative Brain-Computer Interfacing}. In \bibinfo{booktitle}{\emph{Proceedings of the SIGCHI Conference on Human Factors in Computing Systems}} \emph{(\bibinfo{series}{CHI '20})}. \bibinfo{pages}{1–14}.
\newblock
\urldef\tempurl%
\url{https://doi.org/10.1145/3313831.3376288}
\showDOI{\tempurl}


\bibitem[Davis et~al\mbox{.}(2023)]%
        {9987515}
\bibfield{author}{\bibinfo{person}{Keith~M. Davis}, \bibinfo{person}{Michiel Spape}, {and} \bibinfo{person}{Tuukka Ruotsalo}.} \bibinfo{year}{2023}\natexlab{}.
\newblock \showarticletitle{Contradicted by the Brain: Predicting Individual and Group Preferences via Brain-Computer Interfacing}.
\newblock \bibinfo{journal}{\emph{IEEE Trans. Affect. Comput.}} \bibinfo{volume}{14}, \bibinfo{number}{4} (\bibinfo{year}{2023}), \bibinfo{pages}{3094--3105}.
\newblock
\urldef\tempurl%
\url{https://doi.org/10.1109/TAFFC.2022.3225885}
\showDOI{\tempurl}


\bibitem[de~la Torre-Ortiz et~al\mbox{.}(2020)]%
        {de2020brain}
\bibfield{author}{\bibinfo{person}{Carlos de~la Torre-Ortiz}, \bibinfo{person}{Michiel~M. Spap{\'e}}, \bibinfo{person}{Lauri Kangassalo}, {and} \bibinfo{person}{Tuukka Ruotsalo}.} \bibinfo{year}{2020}\natexlab{}.
\newblock \showarticletitle{Brain relevance feedback for interactive image generation}. In \bibinfo{booktitle}{\emph{Proc. Annual ACM Symposium on User Interface Software and Technology}} \emph{(\bibinfo{series}{UIST '20'})}. \bibinfo{pages}{1060--1070}.
\newblock
\urldef\tempurl%
\url{https://doi.org/10.1145/3379337.3415821}
\showDOI{\tempurl}


\bibitem[Kawala-Sterniuk et~al\mbox{.}(2021)]%
        {kawala2021summary}
\bibfield{author}{\bibinfo{person}{Aleksandra Kawala-Sterniuk}, \bibinfo{person}{Natalia Browarska}, \bibinfo{person}{Amir Al-Bakri}, \bibinfo{person}{Mariusz Pelc}, \bibinfo{person}{Jaroslaw Zygarlicki}, \bibinfo{person}{Michaela Sidikova}, \bibinfo{person}{Radek Martinek}, {and} \bibinfo{person}{Edward~Jacek Gorzelanczyk}.} \bibinfo{year}{2021}\natexlab{}.
\newblock \showarticletitle{Summary of over fifty years with brain-computer interfaces—a review}.
\newblock \bibinfo{journal}{\emph{Brain Sciences}} \bibinfo{volume}{11}, \bibinfo{number}{1} (\bibinfo{year}{2021}), \bibinfo{pages}{43}.
\newblock
\urldef\tempurl%
\url{https://doi.org/10.3390/brainsci11010043}
\showDOI{\tempurl}


\bibitem[Kostas et~al\mbox{.}(2021)]%
        {Kostas21fhn}
\bibfield{author}{\bibinfo{person}{Demetres Kostas}, \bibinfo{person}{Stephane Aroca{-}Ouellette}, {and} \bibinfo{person}{Frank Rudzicz}.} \bibinfo{year}{2021}\natexlab{}.
\newblock \showarticletitle{{BENDR:} using transformers and a contrastive self-supervised learning task to learn from massive amounts of {EEG} data}.
\newblock \bibinfo{journal}{\emph{Front. Hum. Neurosci.}}  \bibinfo{volume}{15} (\bibinfo{year}{2021}).
\newblock
\urldef\tempurl%
\url{https://doi.org/10.3389/fnhum.2021.653659}
\showDOI{\tempurl}


\bibitem[Li et~al\mbox{.}(2021)]%
        {9264220}
\bibfield{author}{\bibinfo{person}{Ren Li}, \bibinfo{person}{Jared~S. Johansen}, \bibinfo{person}{Hamad Ahmed}, \bibinfo{person}{Thomas~V. Ilyevsky}, \bibinfo{person}{Ronnie~B. Wilbur}, \bibinfo{person}{Hari~M. Bharadwaj}, {and} \bibinfo{person}{Jeffrey~Mark Siskind}.} \bibinfo{year}{2021}\natexlab{}.
\newblock \showarticletitle{The Perils and Pitfalls of Block Design for {EEG} Classification Experiments}.
\newblock \bibinfo{journal}{\emph{IEEE Trans. Pattern Anal. Mach. Intell.}} \bibinfo{volume}{43}, \bibinfo{number}{1} (\bibinfo{year}{2021}), \bibinfo{pages}{316--333}.
\newblock
\urldef\tempurl%
\url{https://doi.org/10.1109/TPAMI.2020.2973153}
\showDOI{\tempurl}


\bibitem[Mizgajski and Morzy(2019)]%
        {mizgajski2019affective}
\bibfield{author}{\bibinfo{person}{Jan Mizgajski} {and} \bibinfo{person}{Miko{\l}aj Morzy}.} \bibinfo{year}{2019}\natexlab{}.
\newblock \showarticletitle{Affective recommender systems in online news industry: how emotions influence reading choices}.
\newblock \bibinfo{journal}{\emph{User Model. User-Adapt. Interact.}} \bibinfo{volume}{29}, \bibinfo{number}{2} (\bibinfo{year}{2019}), \bibinfo{pages}{345--379}.
\newblock
\urldef\tempurl%
\url{https://doi.org/10.1007/s11257-018-9213-x}
\showDOI{\tempurl}


\bibitem[Moreno-Alcayde et~al\mbox{.}({[n.\,d.]})]%
        {Moreno24paa}
\bibfield{author}{\bibinfo{person}{Yoelvis Moreno-Alcayde}, \bibinfo{person}{Tuukka Ruotsalo}, \bibinfo{person}{Luis~A. Leiva}, {and} \bibinfo{person}{V.~Javier Traver}.} \bibinfo{year}{[n.\,d.]}\natexlab{}.
\newblock \showarticletitle{Affective annotation of videos from {EEG}-based crowdsourcing}.
\newblock  (\bibinfo{year}{[n.\,d.]}).
\newblock
\newblock
\shownote{Under review}.


\bibitem[Moreno-Alcayde et~al\mbox{.}(2024)]%
        {Yoelvis24bmel}
\bibfield{author}{\bibinfo{person}{Yoelvis Moreno-Alcayde}, \bibinfo{person}{V.~Javier Traver}, {and} \bibinfo{person}{Luis~A. Leiva}.} \bibinfo{year}{2024}\natexlab{}.
\newblock \showarticletitle{Sneaky Emotions: Impact of Data Partitions in Affective Computing Experiments with Brain-Computer Interfacing}.
\newblock \bibinfo{journal}{\emph{Biomed. Eng. Lett.}}  \bibinfo{volume}{14} (\bibinfo{year}{2024}), \bibinfo{pages}{103--113}.
\newblock
\urldef\tempurl%
\url{https://doi.org/10.1007/s13534-023-00316-5}
\showDOI{\tempurl}


\bibitem[Pelc et~al\mbox{.}(2023)]%
        {pelc2023machine}
\bibfield{author}{\bibinfo{person}{Mariusz Pelc}, \bibinfo{person}{Dariusz Miko{\l}ajewski}, \bibinfo{person}{Tuukka Ruotsalo}, \bibinfo{person}{Luis~A Leiva}, \bibinfo{person}{Adam Sudo{\l}}, \bibinfo{person}{Edward~Jacek Gorzela{\'n}czyk}, \bibinfo{person}{Adam {\L}ysiak}, {and} \bibinfo{person}{Aleksandra Kawala-Sterniuk}.} \bibinfo{year}{2023}\natexlab{}.
\newblock \showarticletitle{Machine Learning-Based Cascade Filtering System for {fNIRS} Data Analysis}. In \bibinfo{booktitle}{\emph{Proc. of Progress in Applied Electrical Engineering}} \emph{(\bibinfo{series}{PAEE '23})}. \bibinfo{pages}{1--5}.
\newblock
\urldef\tempurl%
\url{https://doi.org/10.1109/PAEE59932.2023.10244522}
\showDOI{\tempurl}


\bibitem[Prange et~al\mbox{.}(2021)]%
        {prange2021investigating}
\bibfield{author}{\bibinfo{person}{Sarah Prange}, \bibinfo{person}{Sven Mayer}, \bibinfo{person}{Maria-Lena Bittl}, \bibinfo{person}{Mariam Hassib}, {and} \bibinfo{person}{Florian Alt}.} \bibinfo{year}{2021}\natexlab{}.
\newblock \showarticletitle{Investigating User Perceptions Towards Wearable Mobile Electromyography}. In \bibinfo{booktitle}{\emph{Proc. IFIP Conf. on Human-Computer Interaction}} \emph{(\bibinfo{series}{INTERACT '21})}. \bibinfo{pages}{339--360}.
\newblock
\urldef\tempurl%
\url{https://doi.org/10.1007/978-3-030-85610-6_20}
\showDOI{\tempurl}


\bibitem[Ruotsalo et~al\mbox{.}(2023a)]%
        {Ruotsalo23tac}
\bibfield{author}{\bibinfo{person}{Tuukka Ruotsalo}, \bibinfo{person}{Kalle Mäkelä}, {and} \bibinfo{person}{Michiel Spapé}.} \bibinfo{year}{2023}\natexlab{a}.
\newblock \showarticletitle{Crowdsourcing Affective Annotations via {fNIRS-BCI}}.
\newblock \bibinfo{journal}{\emph{{IEEE} Transactions on Affective Computing}} (\bibinfo{year}{2023}).
\newblock
\urldef\tempurl%
\url{https://doi.org/10.1109/TAFFC.2023.3273916}
\showDOI{\tempurl}


\bibitem[Ruotsalo et~al\mbox{.}(2023b)]%
        {Ruotsalo23a_affective}
\bibfield{author}{\bibinfo{person}{Tuukka Ruotsalo}, \bibinfo{person}{Kalle Mäkelä}, \bibinfo{person}{Michiel Spapé}, {and} \bibinfo{person}{Luis~A. Leiva}.} \bibinfo{year}{2023}\natexlab{b}.
\newblock \showarticletitle{Affective Relevance: Inferring Emotional Responses via {fNIRS} Neuroimaging}. In \bibinfo{booktitle}{\emph{Proc. Intl. ACM SIGIR Conf. on Research and Development in Information Retrieval}} \emph{(\bibinfo{series}{SIGIR '23})}.
\newblock
\urldef\tempurl%
\url{https://doi.org/10.1145/3539618.3591946}
\showDOI{\tempurl}


\bibitem[Ruotsalo et~al\mbox{.}(2023c)]%
        {Ruotsalo23b_affective}
\bibfield{author}{\bibinfo{person}{Tuukka Ruotsalo}, \bibinfo{person}{Kalle Mäkelä}, \bibinfo{person}{Michiel Spapé}, {and} \bibinfo{person}{Luis~A. Leiva}.} \bibinfo{year}{2023}\natexlab{c}.
\newblock \showarticletitle{Feeling Positive? Predicting Emotional Image Similarity from Brain Signals}. In \bibinfo{booktitle}{\emph{Proceedings of the ACM Intl. Conf. on Multimedia}} \emph{(\bibinfo{series}{MM '23})}.
\newblock
\urldef\tempurl%
\url{https://doi.org/10.1145/3581783.3613442}
\showDOI{\tempurl}


\bibitem[Spapé et~al\mbox{.}(2023a)]%
        {9353984}
\bibfield{author}{\bibinfo{person}{Michiel Spapé}, \bibinfo{person}{Keith~M. Davis}, \bibinfo{person}{Lauri Kangassalo}, \bibinfo{person}{Niklas Ravaja}, \bibinfo{person}{Zania Sovijärvi-Spapé}, {and} \bibinfo{person}{Tuukka Ruotsalo}.} \bibinfo{year}{2023}\natexlab{a}.
\newblock \showarticletitle{Brain-Computer Interface for Generating Personally Attractive Images}.
\newblock \bibinfo{journal}{\emph{IEEE Trans. Affect. Comput.}} \bibinfo{volume}{14}, \bibinfo{number}{1} (\bibinfo{year}{2023}), \bibinfo{pages}{637--649}.
\newblock
\urldef\tempurl%
\url{https://doi.org/10.1109/TAFFC.2021.3059043}
\showDOI{\tempurl}


\bibitem[Spapé et~al\mbox{.}(2023b)]%
        {Spape23tac}
\bibfield{author}{\bibinfo{person}{Michiel Spapé}, \bibinfo{person}{Kalle Mäkelä}, {and} \bibinfo{person}{Tuukka Ruotsalo}.} \bibinfo{year}{2023}\natexlab{b}.
\newblock \showarticletitle{{NEMO}: A Database for Emotion Analysis Using Functional Near-Infrared Spectroscopy}.
\newblock \bibinfo{journal}{\emph{IEEE Trans. Affect. Comput.}} (\bibinfo{year}{2023}).
\newblock
\urldef\tempurl%
\url{https://doi.org/10.1109/TAFFC.2023.3315971}
\showDOI{\tempurl}


\bibitem[Srinivasan(1999)]%
        {Srinivasan99}
\bibfield{author}{\bibinfo{person}{Ramesh Srinivasan}.} \bibinfo{year}{1999}\natexlab{}.
\newblock \showarticletitle{Methods to Improve the Spatial Resolution of {EEG}}.
\newblock \bibinfo{journal}{\emph{International Journal of Bioelectromagnetism}} \bibinfo{volume}{1}, \bibinfo{number}{1} (\bibinfo{year}{1999}), \bibinfo{pages}{102--111}.
\newblock
\urldef\tempurl%
\url{https://doi.org/10.1007/s10484-017-9371-3}
\showDOI{\tempurl}


\bibitem[Wieczorek et~al\mbox{.}(2023)]%
        {wieczorek2023custom}
\bibfield{author}{\bibinfo{person}{Anna Wieczorek}, \bibinfo{person}{Edward~Jacek Gorzela{\'n}czyk}, \bibinfo{person}{Mariusz Pelc}, \bibinfo{person}{Saravanakumar Duraisamy}, \bibinfo{person}{Luis~A Leiva}, {and} \bibinfo{person}{Aleksandra Kawala-Sterniuk}.} \bibinfo{year}{2023}\natexlab{}.
\newblock \showarticletitle{Custom-made Near Infrared Spectroscope as a Tool for Obtaining Information Regarding the Brain Condition}. In \bibinfo{booktitle}{\emph{Proc. Intl. Conf. on Methods and Models in Automation and Robotics}} \emph{(\bibinfo{series}{MMAR '23})}. \bibinfo{pages}{256--263}.
\newblock
\urldef\tempurl%
\url{https://doi.org/10.1109/MMAR58394.2023.10242471}
\showDOI{\tempurl}


\end{thebibliography}

\end{document}